\documentstyle[12pt]{article}
\begin{document}
\title{\Large{{\bf A LINEAR EQUATION FOR WILSON LOOPS}}}

\vspace{1.5cm}

\author{~\\{ Poul Olesen} \\~\\
\it The Niels Bohr Institute\\ \it The Niels Bohr International Academy\\ \it 
Blegdamsvej 17\\ \it DK-2100 Copenhagen {\O}
\\ \it Denmark}
\date{\today}
\maketitle
\vfill
\begin{abstract}
The Makeenko-Migdal loop equation is non-linear and first order in the
area derivative, but we show that for simple 
loops in QCD$_2$ it is possible to reformulate this equation as
a linear equation with  second order derivatives. This equation
is a bound state Schr\"odinger equation with a three dimensional
Coulomb potential. Thus, loop dynamics leads to a surprising new 
picture of 
confinement, where this phenomenon is due to a (bound state) localization
in loop space,
with the Wilson loops decaying exponentially outside a characteristic radius.
\end{abstract}
\vfill

\newpage

The Makeenko-Migdal loop equation \cite{yuri} is an exact consequence of the 
non-linear dynamics in $N=\infty$ QCD, and is therefore also non-linear. It is
of first order in the area derivative. However, although the equation
is closed, loop space is enormous, so it is not easy to gain insight into how
dynamics look like in this space. In the present
letter we restrict ourselves to simple loops with an arbitrary number of 
windings $n$ corresponding to $n$ quarks propagating along the boundary
of the loops.  We then reformulate the loop equation for these
simple loops in  QCD$_2$ starting from the Makeenko-Migdal equation. The
result is an equation which is linear but of second order in the derivatives.
This is obtained from the Hopf equation in QCD$_2$, which can be generalized 
to 
a functional equation in higher dimensions, so the two dimensional result may 
be generalizable to higher dimensions. In any case, our linear equation
represents a new picture of confinement as a bound state in a three 
dimensional space. The interaction between the $n$ quarks is given by
the three dimensional Coulomb potential in loop space! This shows that the
dynamics in loop space is rather surprising and is not in a simple
correspondence with perturbative expansions .

For simple loops with a number $n$ of windings it is possible to derive 
an equation for the spectral density of eigenvalues
\begin{equation}
\rho_C (\theta )=\frac{1}{2\pi}~\left[1+2\sum_{n=1}^{\infty}~W^{(n)}(C)
\cos (n\theta )\right],
\label{1}
\end{equation}
where $W^{(n)}(C)$ is the Wilson loop for the simple curve $C$ winding
$n$ times. The equation is \cite{poul}
\begin{equation}
\partial_\mu\frac{\delta\rho_C(\theta)}{\delta \sigma_{\mu\nu}(x)}=
2g_0^2N\oint_Cdy_\nu~\delta^{(D)}(x-y)~\frac{\partial}{\partial\theta}
\left[\rho_C(\theta)~{\rm P}\int_{-\infty}^\infty d\alpha~\frac{\rho_C(\alpha)}
{\theta-\alpha}\right],
\label{2}
\end{equation}
where P means the principal value of the integral.
If we introduce the resolvent (i.e. the Hilbert transform)
\begin{equation}
f_C(\theta)=\int_{-\infty}^\infty d\alpha~\frac{\rho_C(\alpha)}
{\theta-\alpha}={\rm P}\int_{-\infty}^\infty d\alpha~\frac{\rho_C(\alpha)}
{\theta-\alpha}+i\pi\rho_C(\theta),
\label{3}
\end{equation}
we can rewrite eq. (\ref{2}) as
\begin{equation}
\partial_\mu\frac{\delta f_C(\theta)}{\delta \sigma_{\mu\nu}(x)}=
2g_0^2N\oint_Cdy_\nu~\delta^{(D)}(x-y)~f_C(\theta)\frac{\partial f_C(\theta)}
{\partial\theta}.
\label{4}
\end{equation}
This equation is non-linear, as expected.

In two dimensions eq. (\ref{4}) reduces to the partial differential 
equation \cite{poul},
\begin{equation}
\frac{\partial f(A,\theta)}{\partial A}+f(A,\theta)\frac{\partial f(A,\theta)}
{\partial \theta}=0,
\label{5}
\end{equation}
where it was used that in $D=2$ the Wilson loops depend only on the area
$A$, which is measured in units of $g_0^2N$.
Eq. (\ref{5}) is called the Hopf (or Burgers\footnote{Although the
viscosity term usualy occuring in Burgers eq. is absent here.}) equation. In 
\cite{gross} it is shown that this type
of equation, or a slight generalization, occurs for any multiplicatively
free family of random variables \cite{free}. Eq. (\ref{4}) is valid in all
dimensions and is a Hopf equation in loop space. 

We shall now show that for the boundary conditions $W^{(n)}(0)=1$ 
the non-linear first order Hopf equation (\ref{5}) reduces to a
linear second order equation. Let us use the notation $f(0,\theta)=
f_0(\theta)$. By means of the method of characteristics 
eq. (\ref{5}) can be reformulated to
\begin{equation}
f(A,\theta)=f_0(\xi (A,\theta) ),~~\xi (A,\theta)=\theta-Af_0(\xi(A,\theta)),
\label{6}
\end{equation}
and the various derivatives can be expressed as
\begin{eqnarray}
&& \frac{\partial f(A,\theta)}{\partial A}=\frac{-f_0(\xi)f_0'(\xi)}
{1+f_0'(\xi)A},~~\frac{\partial f(A,\theta)}{\partial \theta}=
\frac{f_0'(\xi)}{1+f_0'(\xi)A},\nonumber \\
&&\frac{\partial \xi}{\partial A}=\frac{-f_0(\xi)}{1+f_0'(\xi)A},~~~~~~~~~~~
\frac{\partial \xi}{\partial \theta}=\frac{1}{1+f_0'(\xi)A}.
\label{7}
\end{eqnarray}
Here the prime means differentiation with respect to $\xi$. From
(\ref{7}) we obtain
\begin{equation}
\frac{\partial^2 Af(A,\theta)}{\partial A^2}=
\frac{1}{(1+Af_0'(\xi))^3}\left(-2f_0(\xi)f_0'(\xi)+A(-2f_0(\xi)(f_0'(\xi))^2+
f_0(\xi)^2f_0''(\xi))\right),
\label{8}
\end{equation}
and
\begin{equation}
\frac{\partial^2f(A,\theta)}{\partial\theta^2}=\frac{f_0''(\xi)}
{(1+Af_0'(\xi))^3}.
\label{9}
\end{equation}
By comparison of eqs. (\ref{8}) and (\ref{9}) we see that we can form
a linear second order differential equation of the form
\begin{equation}
\frac{\partial^2 Af(A,\theta)}{\partial A^2}=(a+bA)~\frac{\partial^2
f(A,\theta)}{\partial\theta^2},
\label{10}
\end{equation}
with $a$ and $b$ to be determined, provided
\begin{equation}
af_0''(\xi)=-2f_0(\xi)f_0'(\xi),
\label{11}
\end{equation}
and
\begin{equation}
bf_0''(\xi)=-2f_0(\xi)(f_0'(\xi))^2+f_0(\xi)^2f_0''(\xi).
\label{12}
\end{equation}
To solve, insert eq. (\ref{11}) in (\ref{12}), which gives 
\begin{equation}
f_0(\xi(A,\theta))=f(A,\theta )=\sqrt{b}~\coth (\sqrt{b}~\xi(A,\theta) /a).
\label{13}
\end{equation}
By inserting this result we see that eqs. (\ref{11}) and (\ref{12}) are 
satisfied without any restrictions on $a$ and $b$.

It now remains to see whether the result (\ref{13}) fits with QCD$_2$,
where 
\begin{equation}
\rho_{A=0}(\theta)=\sum_{n=-\infty}^\infty ~\delta (\theta-2\pi n),
\label{14}
\end{equation}
so from eq. (\ref{3}) we get
\begin{equation}
f(A=0,\theta)=f_0(\theta)=\sum_{n=-\infty}^\infty \frac{1}{\theta-2\pi n}=
\frac{1}{2}~\cot \frac{\theta}{2}.
\label{15}
\end{equation}
If we compare this with the solution (\ref{13}) we see that QCD$_2$ satisfies
the linear second order differential equation (\ref{10}) provided
$a=1$ and $\sqrt{b}=i/2$. Hence the differential equation (\ref{10})
becomes
\begin{equation}
\frac{\partial^2 Af(A,\theta)}{\partial A^2}-\frac{1}{4}\left(\frac{4}{A}
-1\right)
\frac{\partial^2 Af(A,\theta)}{\partial \theta^2}=0.
\label{16}
\end{equation}
This linear equation can thus replace the non-linear Hopf equation (\ref{5})
when the QCD$_2$ initial value (\ref{15}) is used. 

It should be mentioned that in a similar manner one can produce other
second order linear versions of the Hopf equation. For example, instead of
eq. (\ref{16}) we could study
\begin{equation}
\frac{\partial^2 f(A,\theta)}{\partial A^2}=(c+dA)\frac{\partial^2 
f(A,\theta)}{\partial \theta^2},
\label{17}
\end{equation}
which leads to a cubic equation for the initial value function
\begin{equation}
cf_0(\xi)-1/3~f_0(\xi)^3=d\xi+{\rm const.}
\label{18}
\end{equation}
This equation has no relation with QCD$_2$, and the linear differential 
equation (\ref{17}) must therefore be rejected. Only eqs. (\ref{10}) and
(\ref{16}) are relevant.

As shown in ref. \cite{gross} the Hopf equation (or simple generalizations)
occurs in any multiplicatively free family of random variables. Thus the
method used above to rewrite the Hopf equation in terms of  equations which 
incorporate the initial conditions can be used to obtain a linear
replacement of the relevant Hopf equation. The point is then to 
chose the linear equation such that the initial function is physically
relevant, which may not always be possible.

Let us now return to the linear QCD$_2$ equation (\ref{16}). Suppose this
is all we know, then how do we solve it? The simplest way is to use 
the factorization $f(A,\theta)=f_1(A)f_2(\theta)$. Eq. (\ref{16}) can be 
separated in the two variables, and the resulting equation for $f_2(\theta)$ 
is trivial and has the solution $ f_2(\theta)=\exp (\lambda\theta)$, where 
$\lambda^2$ is the separation constant. Since $f(A,\theta)$ is defined to
be periodic in $\theta$ with period $2\pi$, it follows that $\lambda$ should 
be an integer times $i$. For each such $\lambda$ the function $f_1$ should be 
proportional to a $W^{(n)}(A)$. Referring to 
of eqs. (\ref{1}) and (\ref{3}) the separation method gives
\begin{equation}
f(A,\theta)=i\left(\frac{1}{2}+\sum_{n=1}^\infty W^{(n)}(A)e^{-in\theta}
\right),
\label{19}
\end{equation}
and the Wilson loops should satisfy the separation equation
\begin{equation}
\frac{d^2W^{(n)}}{dA^2}+\frac{2}{A}~\frac{dW^{(n)}}{dA}+\frac{n^2}{4}
\left(\frac{4}{A}-1\right)W^{(n)}=0.
\label{20}
\end{equation}
This equation has the solution
\begin{equation}
W^{(n)}(A)=\frac{1}{n}L_{n-1}^{1}(nA)e^{-nA/2},
\label{21}
\end{equation}
where the $L'$s are the Laguerre polynomials of type 1, and where
we used the boundary condition $W^{(n)}(0)=1$ to fix an arbitrary constant. 
Eq. (\ref{21})  is the well known solution for winding loops 
\cite{kostov} in QCD$_2$.  

Eq. (\ref{20}) can of course be derived directly from the known 
solution (\ref{21}).
The reason we have started from the Hopf equation (\ref{5}) is that
it can be generalized to higher dimensions as is seen  in
eq. (\ref{4}). In $ D=3$ or 4 it is not true that the resolvent
$f_C(\theta)$ only depends on the area, even for simple loops. However, one
may speculate that area dependence sets in asymptotically for sufficiently 
large loops. On the other hand, from the works in refs. \cite{poly} it
follows that the loop equation in $D=4$ has features quite different from its
$D=2$ counterpart.  It would be an interesting 
project for future research to try to invent a  generalization of the
method of characteristics (leading to eqs. (\ref{6}) and (\ref{7}) in
two dimensions)
to the functional version of the Hopf equation valid in higher than two 
dimensions, incorporating the complexities discussed in refs. \cite{poly}. 

In a recent work by Narayanan and Neuberger \cite{herbert} they
have considered the spectral density in four dimensions and found that
it is similar to the density in two dimensions. This approach refers to
Wilson loops in which the Coulomb-like contribution has been smeared in order
to avoid divergences, thereby making the $D=4$ case similar to the $D=2$ case.
If the the functional Hopf equation (\ref{4}) could be similarly
smeared it would probably be 
similar to its two dimensional version for large {\it smeared} loops. 

Eq. (\ref{20}) is the
(zero angular momentum) radial Schr\"odinger equation for the hydrogen
atom if we identify the radial variable with the loop quantity $A$ and the  
charge with $n$.
Since $n$ quarks are running around on the boundary, this is a
quite natural identification. Similarly the ``bound state'' energy is $-n^2/4$.
The point $A=4$ is essentialy the ``Bohr radius'' of the atom. Inside this 
radius the ``wave function'' $W^{(n)}$ oscillates
as a function of $A$, whereas outside this
quantity is exponentially damped. Although confinement is often considered
trivial in two dimensions, the ``wave functions'' $W^{(n)}$ behave in a
rather non-trivial manner, especially for larger $n'$s. Similar behavior
does not occur in the Abelean case in two dimensions. Since there exist the 
hope that the Hopf equation can be treated by a functional
version of the method of characteristics in three or four dimensions, 
the bound state picture of confinement may be valid also in higher dimensions.

Loop space dynamics thus results in a rather surprising 
picture of confinement, where the area acts as a radial variable in 
a three dimensional space in the sense that the force between the quarks
with charges $n$ is governed by the three dimensional Coulomb potential.
The winding Wilson loops $W^{(n)}(A)$ act as wave functions satisfying a
Schr\"odinger equation. Inside the localization radius $A=4$ these wave
functions oscillate (more and more when $n$ increases) and outside
$A=4$ they decay exponentially.
These three dimensions are genuinely a product of the loop dynamics,
which seems to have forgotten about the  two dimensions where it lives. It
would be interesting to investigate other more complicated loops to see
to what extent this picture of confinement is universal. 
 
One can speculate that the reason for the {\it localized} bound state
picture is that QCD$_2$ basically is a family of multiplicatively free
random variables \cite{gross}, which is in analogy with Anderson localization 
of the wave function in the presence of random fields in quantum mechanics. 
It also could well be that a similar phenomenon occurs in higher dimensions 
due to the expected random fields in the QCD vacuum.


\begin{thebibliography}{X}

\bibitem{yuri}  Yu. Makeenko and A. A. Migdal, Physics Letters  88B (1979)
135

\bibitem{poul} B. Durhuus and P. Olesen, Nucl. Phys. B184 (1981) 461; B. 
Durhuus and P. Olesen, Nucl. Phys. B184 (1981) 135

\bibitem{gross} R. Gopakumar and D. J. Gross, Nucl. Phys. B451 (1995) 379,
{\tt arXiv:hep-th/9411021} 

\bibitem{free} D. V. Voiculescu, K. J. Dykema and A. Nica, {\it Free Random 
Variables}, AMS, Providence (1992)

\bibitem{kostov} V. A. Kazakov and I. K. Kostov, Nucl. Phys. B176 (1980) 199;
P. Rossi, Annals Phys. 132 (1981) 463

\bibitem{poly} A. M. Polyakov and V. Rychkov, Nucl. Phys. B581 (2000) 116; A. 
M. Polyakov and V. Rychkov, Nucl. Phys. B594 (2001) 272; M. Bochicchio, JHEP 
0709 (2007) 033

\bibitem{herbert} R. Narayanan and H. Neuberger, JHEP 03 (2006) 064


\end{thebibliography}
\end{document}